\documentclass[runningheads]{llncs}

\usepackage[T1]{fontenc}
\usepackage{graphicx,verbatim}
\usepackage{graphicx}
\usepackage{amssymb}
\usepackage{booktabs}
\usepackage{multirow} 
\usepackage{amssymb}
\usepackage{enumitem}
\usepackage{pifont}
\usepackage{xcolor} 
\usepackage{amsmath}
\usepackage{wrapfig} 
\usepackage{colortbl}
\usepackage{algorithm}
\usepackage{algpseudocode}

\begin{document}

\title{Unsupervised Cardiac Video Translation Via Motion Feature Guided Diffusion Model}

\begin{comment}  %% Removed for anonymized MICCAI 2025 submission

\end{comment}

\author{Swakshar Deb\inst{1} \and
 Nian Wu\inst{1} \and
Frederick H. Epstein\inst{2} \and Miaomiao Zhang\inst{1,3}}

\authorrunning{S. Deb et al.}

\institute{Department of Electrical and Computer Engineering, University of Virginia, USA \and
Department of Biomedical Engineering, University of Virginia, USA \and 
Department of Computer Science, University of Virginia, USA}
    
\maketitle              % typeset the header of the contribution
\begin{abstract}
This paper presents a novel motion feature guided diffusion model for unpaired video-to-video translation (MFD-V2V), designed to synthesize dynamic, high-contrast cine cardiac magnetic resonance (CMR) from lower-contrast, artifact-prone displacement encoding with stimulated echoes (DENSE) CMR sequences. To achieve this, we first introduce a Latent Temporal Multi-Attention (LTMA) registration network that effectively learns more accurate and consistent cardiac motions from cine CMR image videos. A multi-level motion feature guided diffusion model, equipped with a specialized Spatio-Temporal Motion Encoder (STME) to extract fine-grained motion conditioning, is then developed to improve synthesis quality and fidelity. We evaluate our method, MFD-V2V, on a comprehensive cardiac dataset, demonstrating superior performance over the state-of-the-art in both quantitative metrics and qualitative assessments. Furthermore, we show the benefits of our synthesized cine CMRs improving downstream clinical and analytical tasks, underscoring the broader impact of our approach. Our code is publicly available at https://github.com/SwaksharDeb/MFD-V2V.
\end{abstract}

\section{Introduction}

Cardiac Magnetic Resonance (CMR) imaging plays an important role in assessing myocardial strain, which is a key indicator of cardiac dysfunction~\cite{chadalavada2024myocardial}. Among various CMR techniques, displacement encoding with stimulated echoes (DENSE) CMR  has demonstrated superior performance in capturing myocardial motion for highly accurate and reliable strain measurement~\cite{sillanmaki2023measuring}. In contrast to standard cine CMR, which relies on balanced steady-state free precession sequences optimized for high signal-to-noise ratio (SNR) and strong tissue contrast, DENSE CMR encodes myocardial displacement and deformation using stimulated echoes, allowing for high-resolution regional and segmental strain analysis. However, this process comes at the cost of lower SNR, as stimulated echoes retain only a fraction of the original magnetization, leading to reduced signal intensity and noisier magnitude images~\cite{aletras1999dense}. Despite that DENSE CMR excels in providing highly detailed quantitative myocardial motion/strain data, its lower image quality presents challenges for tasks such as segmentation, feature extraction, and texture-based analysis, limiting its applicability in certain clinical applications~\cite{gilliam2012automated}.

Recent advancements in deep learning and generative models offer a promising new approach to synthesizing high-quality cine CMRs from DENSE CMRs, which remains unexplored in the literature. Early research on video-to-video (V2V) translation using generative models, such as generative adversarial networks (GANs), has demonstrated their ability to transform video sequences across domains~\cite{bansal2018recycle,chen2020generative}. However, GAN-based methods often suffer from limited diversity in their translations and are prone to mode collapse, restricting the variability of synthesized outputs~\cite{metz2016unrolled}. More recently, diffusion-based generative models have emerged as a powerful alternative, offering superior diversity, fidelity, and stability over GANs~\cite{rombach2022high}. Despite their advantages, most diffusion-based V2V models ~\cite{hu2023videocontrolnet,liang2024flowvid,chu2024medm,wu2023tune} rely on a paired (supervised) training scenarios, where aligned source-target video pairs are required. However, cine and DENSE CMRs do not naturally form a paired dataset, as they are acquired using fundamentally different imaging sequences, leading to significant disparities in spatial and temporal resolution. 

%In this paper, we introduce a novel motion feature guided diffusion model for unpaired cardiac MRI V2V translation, termed MFD-V2V. In particular, we first propose a latent temporal multihead attention (LTMA) based registration network to effectively capture temporal motion from cine CMR sequences.  which are utilized to guide the diffusion process during training. During inference, we leverage the displacement field provided by the Dense CMR to synthesize paired Cine CMR. Morevoer, we propose a spatio temporal motion encoder (STME) to capture fine grained motion conditioning. 
In this paper, we introduce a novel motion feature guided diffusion model for unpaired cardiac MRI V2V translation, termed MFD-V2V. Our method harnesses the spatiotemporal motion information inherent in cine CMR videos, using learned motion features as conditions to guide the generative diffusion model in learning the complex distribution of cine CMRs. During inference, we leverage the displacement motion field provided by DENSE-CMR to generate realistic and temporally consistent cine CMR sequences; hence bridging the gap in unpaired cardiac MRI translation with improved anatomical and motion fidelity. Our contributions are threefold:
\begin{enumerate}[label=(\roman*)]
\item  We are the first to develop a generative diffusion V2V model to synthesize high quality cine CMR from DENSE CMR sequences.
\item Introduce a latent temporal multihead attention (LTMA) based registration network to effectively learn spatiotemporal motion from cardiac video sequences. 
\item Develop a generative video diffusion model conditioned on multi-level motion features extracted by a specialized spatiotemporal motion encoder (STME) to enhance synthesis quality and fidelity.
\end{enumerate}
We validate MFD-V2V on cardiac MR images collected from multiple sites~\cite{ghadimi2024deep,lei2025improved}. Experimental results demonstrate that MFD-V2V surpasses existing methods~\cite{ho2022video,zhang2023adding,bansal2018recycle,zhu2017unpaired,chen2020reusing} by generating more realistic and temporally coherent cardiac MR videos. Furthermore, we highlight the advantages of our synthesized data in improving performance on a downstream CMR myocardium segmentation task.

\section{Background: Video Diffusion Model}
\label{sec:background}
This section briefly reviews the concept of video diffusion model (VDM)~\cite{ho2022video}, which serves as a key foundation for our proposed video translation model.

Given a image $x_0$ sampled from the real data distribution $q(x)$, the forward process of the diffusion model is defined as a Markov process in which Gaussian noise is gradually added to $x_0$ following a variance scheduler $\{\beta_t \in (0,1)\}_{t=1}^T$. This process is formulated as
$$q(x_{1:T}|x_0) = \prod_{t=1}^T q(x_t|x_{t-1}), \text{where} \, \,   q(x_t|x_{t-1})=\mathcal{N}(x_t, \sqrt{1-\beta_t}x_{t-1}, \beta_t \mathbf{I}).$$
Using the notation $\alpha_t = 1-\beta_t$ and $\bar{\alpha}_t = \prod_{s=1}^t \alpha_s$, the direct formulation of the noising process from $x_0$ to $x_t$ is $q(x_t|x_0) = \mathcal{N}(x_t; \sqrt{\bar\alpha_t}x_0, (1-\bar{\alpha}_t)\mathbf{I})$. Similarly, in reverse process we model the joint distribution, $p_{\theta}(x_{0:T})$, as a markov chain starting from $p(x_{T}) = \mathcal{N}(x_{T}; 0,\mathbf{I})$:
$$p_{\theta}(x_{0:T}) = p(x_T) \prod_{t=1}^T p_{\theta}(x_{t-1}|x_t), \text{where} \, \, p_{\theta}(x_{t-1}|x_t) = \mathcal{N}(x_{t-1}; \mathbf{\mu}_{\theta}(x_t,t),\mathbf{\Sigma}_{\theta}(x_t,t)) $$
where, $\mathbf{\mu}_{\theta}$ and $\mathbf{\Sigma}_{\theta}$ is the predicted mean and variance respectively. Instead of learning the mean $\mathbf{\mu}_{\theta}$, following the prior work \cite{ho2020denoising}, we can parameterize $\mathbf{\mu}_{\theta}$ to a noise predicting network $\epsilon_{\theta}$, and set the variance at the identity. The model is trained by minimizing the following objective, $L_{\epsilon} = \mathbb{E}_{x_{t},\epsilon \sim \mathcal{N}(0,I)}[||\epsilon - \epsilon_{\theta}(x_{t},t) ||^2]$. 

To extend this image-based diffusion framework to videos, the VDM~\cite{ho2022video} employs a 3D U-Net~\cite{cciccek20163d} to factorize over both spatial and temporal dimensions by replacing standard 2D convolutions with 3D convolutions. Additionally, VDM incorporates spatial and temporal self-attention blocks to effectively capture structural details and motion dynamics across frames.

\section{Our Method: MFD-V2V}
\label{sec:methodology}
\begin{figure}[!t]
\includegraphics[width=\textwidth, trim={0.5cm 0.4cm 1.0cm 0.4cm}, clip]{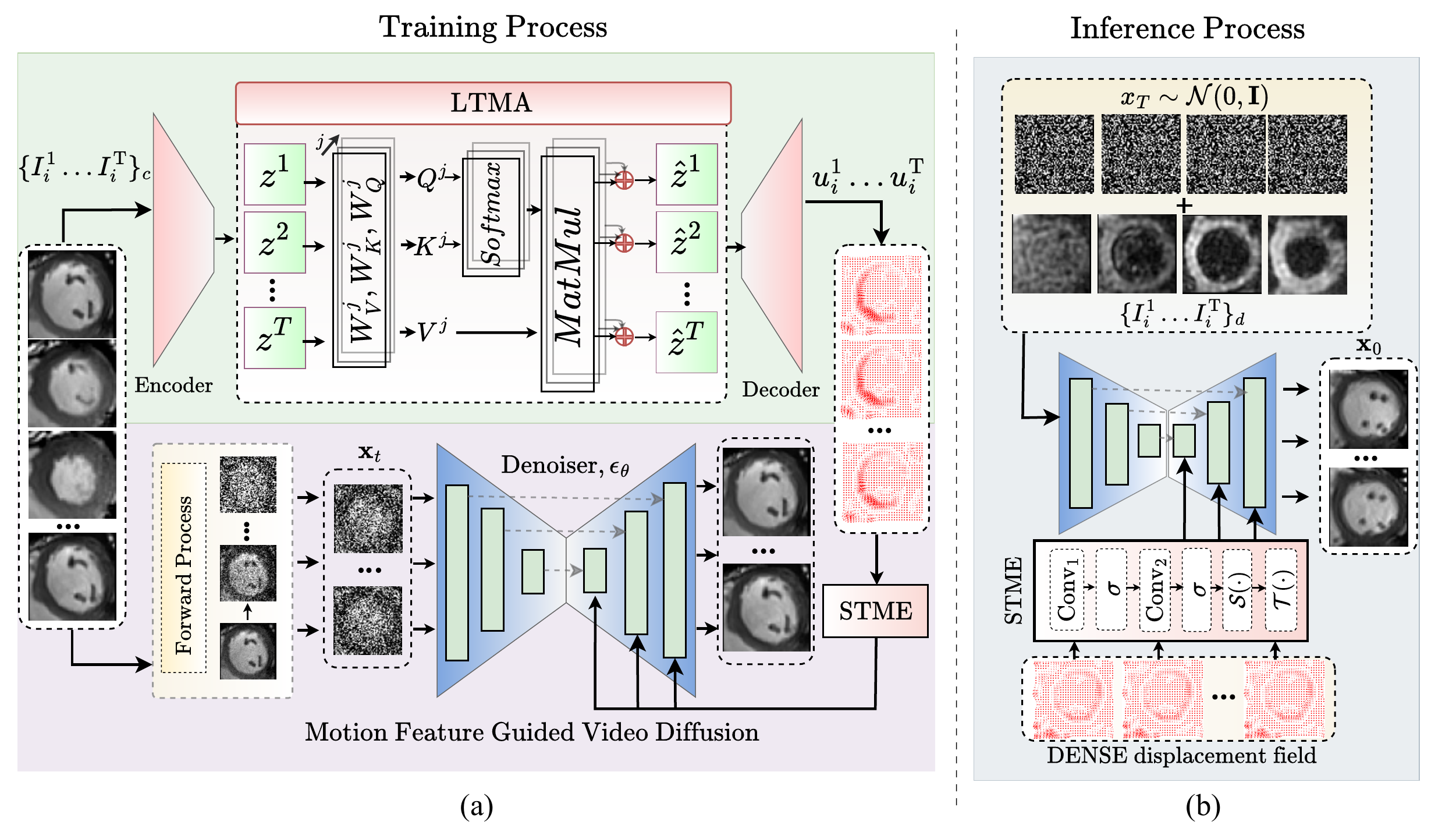}
\caption{An overview of our proposed MFD-V2V model. Left to right: (a) the training process with LTMA registration and multli-level motion feature guided video diffusion. (b) the inference process, where encoded DENSE motion features are used as condition.}  
\label{fig:overall_diagram}
\vspace{-0.4cm}
\end{figure}

This section introduces MFD-V2V, an unsupervised video translation model that for the first time synthesizes high-quality cine CMR from DENSE-CMR video sequences via a motion-guided diffusion model. Our method consists of two main components: (i) a latent temporal multihead attention  registration network, LTMA, to learn temporally consistent and continuous motion from CMR video
sequences; and (ii) a generative video diffusion model conditioned on multi-level motion features extracted by a spatiotemporal motion encoder, STME. An overview of the architecture is illustrated in Fig.~\ref{fig:overall_diagram}. 

\subsection{LTMA Registration Network}
Given a training dataset of $N$ video sequences, where each sequence includes $T+1$ time frames, denoted as $\{I^\tau_i\}, i \in \{1, \cdots, N\}, \tau \in \{0, \cdots, T\}$. That is to say, for the $i$th training data, we have a sequence of image frames $\{I^0_i, \cdots, I^T_i$\}. By setting the first frame $\{I^0_i\}$ as a reference (source) image, there exists a number of $T$ pairwise images, $\{(I_i^0, I_i^1), (I_i^0, I_i^2) \cdots, (I_i^0, I_i^T)\}$, to be aligned by their associated motion/displacement field $\{u_i^1, u_i^2 \cdots, u_i^T\}$. Similar to~\cite{balakrishnan2019voxelmorph,wu2024tlrn}, we employ U-Net architecture as the backbone of our registration encoder, $\mathcal{E}_{\theta_v}$, and decoder, $\mathcal{D}_{\theta_v}$, parameterized by $\theta_v$. The encoder $\mathcal{E}_{\theta_v}$ projects the input image sequences into a latent velocity space $\mathbf{Z}_i = [z_i^1,z_i^2,\ldots,z_i^T] \in \mathbb{R}^{T \times H \times W \times C}$. Here $C$ is the number of feature channels, and $H, W$ represent the height and weight of the encoded motion, respectively. The decoder $\mathcal{D}_{\theta_{v}}$ is then used to project the latent features back to the input image space. 
 
\paragraph*{\bf Motion learning via LTMA.} In contrast to existing methods~\cite{wu2024tlrn,reinhardt2008registration} that rely on recurrent residual networks or long short-term memory to model temporal motions sequentially, we leverage the self-attention mechanism to capture long-range temporal dependencies in the encoded latent space of video motion. This enables a more expressive, scalable, and computationally efficient representation of cardiac dynamics by directly modeling global temporal interactions across all time frames. 

For each encoded motion feature $\mathbf{Z}_i$, we define $\mathbf{Q}_i^{(j)}$, $\mathbf{K}_i^{(j)}$, $\mathbf{V}_i^{(j)}$ as the Query, Key, and Value matrices for the $j$-th attention head. Following similar principles of~\cite{vaswani2017attention}, the output feature $\hat{\mathbf{Z}}_i$ of our proposed LTMA module is formulated as
\begin{align}
    \mathbf{Q}_i^{(j)} &= \mathbf{W}^{(j)}_Q \mathbf{Z}_i,  \mathbf{K}_i^{(j)} = \mathbf{W}^{(j)}_k \mathbf{Z}_i,    \mathbf{V}_i^{(j)} = \mathbf{W}^{(j)}_V \mathbf{Z}_i, \label{eqn:att_1} \\ \nonumber
    \mathbf{\hat{Z}}_i &= \oplus_{j=1}^h \text{Softmax}\big(\mathbf{Q}_i^{(j)}\mathbf{K}_i^{(j)T}\big)\mathbf{V}_i^{(j)}, \label{eqn:att_2}
\end{align}
where $\oplus$ denotes the concatenation operation, $h$ is the total number of attention head and $\mathbf{W}^{(j)}_Q$, $\mathbf{W}^{(j)}_K$, $\mathbf{W}^{(j)}_V$ are the linear projection matrix associated with the $j$-th attention head. \\

\noindent {\bf Network loss.} By defining $\Theta$ for all network parameters, we finally formulate the loss function of LTMA registration network as
\begin{align}
    l(\theta) = \sum_{i=1}^N \sum_{t=1}^T \lambda || I_i^{1} \circ \phi_{i}^{t}(v_i^t(\hat{z}_i^{t}); \Theta)-I_i^t||_2^2 + ||\nabla v_{i}^t(\hat{z}_{i}^t;\Theta) ||^2 + \text{reg}(\Theta),
\end{align}
where $\phi_{i}^{t}$ represents the transformation fields between pairwise frames, parameterized by a stationary velocity field $v_{i}^{t}$ over time~\cite{vercauteren2008symmetric}. For further details, please refer to~\cite{vercauteren2008symmetric}. The motion or displacement field can then be computed as $u_{i}^{t} = \phi_{i}^{t} - id$, where $id$ denotes the identity mapping (i.e., the original image grid). The $\lambda$ is a positive weighting parameter, and $\text{reg}(\cdot)$ denotes the regularization function applied to the network.

\subsection{Multi-Level Motion Feature Guided Video Diffusion}
Given the learned motion fields, $ \mathbf{u}_i = \{u_i^t\}_{t=1}^T$, from our previously introduced LTMA registration network, we now present our multi-level motion feature-guided video diffusion model for synthesizing cine CMR from DENSE CMRs. By conditioning on motion, our goal is to ensure that the generated video not only maintains visual fidelity but also accurately captures the natural dynamic motion of the heart over time.

Since temporal motion fields often contain complex dynamic patterns, instead of directly using them as a condition, we propose leveraging multi-level motion features that capture both coarse and fine-grained aspects of motion. These features better guide the diffusion model, enabling it to generate more realistic and temporally consistent outputs. Given an input motion sequence $\mathbf{u}_i$, we define the overall output of the STME block, $\mathbf{F} \in \mathbb{R}^{T \times H \times W \times C}$, as
 \begin{align}
  \mathbf{F} = \mathcal{T} \circ \mathcal{S}\circ \sigma \circ \text{Conv}_2 \circ \sigma\circ\text{Conv}_1(\mathbf{u}_i), \label{eqn:stme} 
 \end{align}
where, $\circ$ is the function composition. Intuitively, in Eq.~\eqref{eqn:stme}, we first apply two consecutive 3D convolution layers with nonlinear ReLU activations ($\sigma$) to the input motion sequence $\mathbf{u}_i$. These feature maps are then passed through spatial and temporal attention layers~\cite{ho2022video}, $\mathcal{S}(\cdot)$ and $\mathcal{T}(\cdot)$, respectively, to extract fine-grained spatiotemporal features of the motion pattern.

For motion conditioning, we employ cross-attention ~\cite{xu2024magicanimate,zhang2023adding} in the decoder of the denoiser $\epsilon_{\theta}$ (as shown in Fig.~\ref{fig:overall_diagram}(b)), where the learned feature map $\mathbf{F}$ is treated as key and value, while the latent layer representation serves as the query. Finally, the overall training objective is: $L_{\epsilon} = \mathbb{E}_{x_{t},\epsilon \sim \mathcal{N}(0,I)}[||\epsilon - \epsilon_{\theta}(x_{t},t, \mathbf{F}) ||^2]$, where $\epsilon$ is the added Gaussian noise over the input data.

\noindent\textbf{Inference Process:} During inference, we initialize with random Gaussian noise, \( x_T \sim \mathcal{N}(0,\mathbf{I}) \), and incorporate the DENSE magnitude information, \( \{I^1_i, \dots, I^T_i\}_d \). The displacement field, directly provided with the DENSE CMR sequences \cite{ghadimi2024deep}, is used to extract motion conditions (see Fig.~\ref{fig:overall_diagram}(b)). We employ the standard DDPM sampler \cite{ho2020denoising} to iteratively synthesize Cine CMR.

\section{Experimental Evaluation}
We evaluate the effectiveness of our proposed model, MFD-V2V, on cardiac CMR video sequences by comparing it against five state-of-the-art (SOTA) generative models: CycleGAN~\cite{zhu2017unpaired}, NiceGAN~\cite{chen2020reusing}, RecycleGAN~\cite{bansal2018recycle}, VDM~\cite{ho2022video}, and ControlNet~\cite{zhang2023adding}. The VDM model is trained without conditioning, while ControlNet is conditioned on the displacement field provided by LTMA. Since ControlNet is a fine-tuning approach, we use VDM as its backbone (from Tab.~\ref{tab:overall_table}) and train only the ControlNet component, following the method in \cite{zhang2023adding}.

\noindent\textbf{Dataset.} We utilize $741$ cine and DENSE CMR videos of the left ventricular (LV) myocardium collected from $284$ subject, including 124 healthy volunteers and 160 patients with various types of heart disease~\cite{ghadimi2024deep,lei2025improved}. All cine and DENSE CMR sequences were temporally and spatially aligned for efficient network training. In particular, the standard cine sequences were temporally resampled to $40$ frames to match the DENSE temporal resolution. All images were resampled to a \(1 \text{ mm}^2\) resolution and cropped to \(128 \times 128\). Both Cine and DENSE sequences have their corresponding LV contour masks manually annotated. 

\begin{figure*}[!b]
\includegraphics[width=\linewidth, trim={0.9cm 0.4cm 0.7cm 0.3cm}, clip]{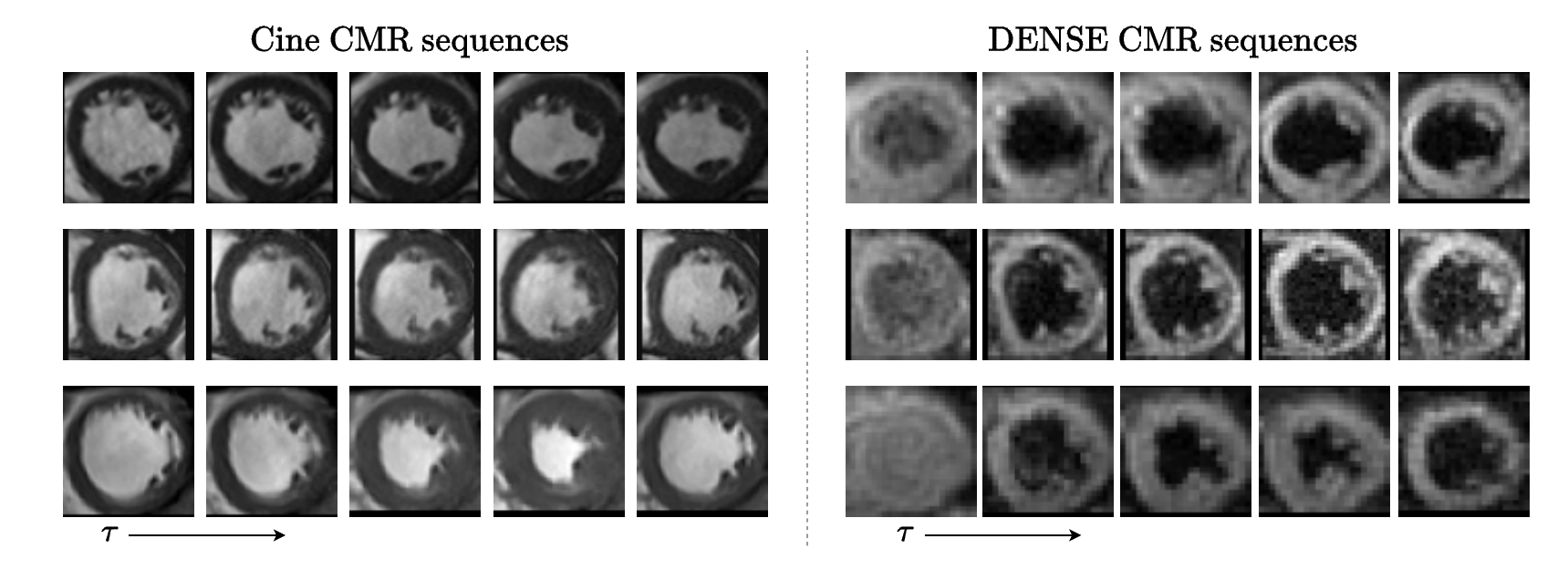}
\caption{Unpaired Cine and DENSE cardiac sequences of the left ventricle from the Cine-DENSE dataset.}
\label{fig:dataset}
\end{figure*}

\subsection{Experimental Design and Implementation Details}
We adopt the 3D-UNet from the Ho \textit{et. al.} \cite{ho2022video} as our denoiser. The model is trained on $10$ frames of $64 \times 64$ spatial resolution focusing on LV. For the training of the denoiser, we use the Adam optimizer with a learning rate of $10^{-5}$, and a batch size of $20$ for $25,000$ training steps. We use the same sigmoid-beta noise scheduler \cite{jabri2023scalable} in the forward process. Following DDPM \cite{ho2020denoising}, we set the total number of diffusion steps as $1000$. The whole framework is implemented using PyTorch and trained on 4 NVIDIA A100 GPUs.

\noindent\textbf{Evaluation Metric.} Since Cine and DENSE do not come with exact pair, this makes direct image-by-image comparison (e.g., PSNR, SSIM, etc) challenging. Hence, instead of direct comparison, we sought to evaluate our model by measuring the distributional shift between the synthesized and original Cine CMR sequences with metrics: Fréchet Inception Distance (FID) \cite{heusel2017gans}, Kernel Inception Distance (KID) \cite{binkowski2018demystifying}, Fréchet Video Distance (FVD) \cite{unterthiner2018towards}, and FID-VID \cite{balaji2019conditional}.

\subsection{Results}

\begin{table}[t]
\centering
\setlength{\tabcolsep}{4pt} % Adjust column spacing (default is 6pt)
\caption{Quantitative analysis on Cine-DENSE cardiac dataset. The best and second best performances are highlighted with bold and underlined, respectively. Here, the Reg. Net. represents registration network.}
\resizebox{\textwidth}{!}{
\begin{tabular}{l l c ccccc}
\toprule
Category & Method & Condition & Reg. Net. & FID $\downarrow$ & KID $\downarrow$ & FVD $\downarrow$ & FID-VID $\downarrow$ \\
\midrule
\multirow{3}{*}{GAN} 
& CycleGAN \cite{zhu2017unpaired} & $\times$ & $\times$ & $135.25$ & $0.3071$ & $141.71$ & $68.567$ \\
& NiceGAN \cite{chen2020reusing} & $\times$ & $\times$ & $122.43$ & $0.2081$ & $131.77$ & $77.013$ \\
& RecyleGAN \cite{bansal2018recycle} & $\times$ & $\times$ & $97.996$ & $0.1591$ & $126.11$ & $68.214$ \\
\midrule
\multirow{3}{*}{DMs} 
& VDM \cite{ho2022video} & $\times$ & $\times$ & $\underline{57.235}$ & $0.0483$ & $75.487$ & $30.148$ \\
& ControlNet \cite{zhang2023adding} & Motion & LTMA & $61.452$ & $\underline{0.0449}$ & $\underline{70.217}$ & $\underline{27.479}$ \\

\rowcolor{pink!30}
& Ours & Motion & LTMA & $\boldsymbol{43.432}$ & $\boldsymbol{0.0179}$ & $\boldsymbol{50.962}$ & $\boldsymbol{20.124}$ \\
\bottomrule
\end{tabular}
}
\label{tab:overall_table}
\end{table}

\begin{table}[t]
\centering
\setlength{\tabcolsep}{5pt} % Adjust column spacing (default is 6pt)
\caption{Ablation study on Cine-DENSE dataset. We incrementally add or replace component to compare their contribution.}
\begin{tabular}{c c ccccc}
\toprule
STME & LTMA & FID $\downarrow$ & KID $\downarrow$ & FVD $\downarrow$ & FID-VID $\downarrow$ \\
\midrule
$\times$  & $\times$ & $55.103$ & $0.0412$ & $69.266$ & $30.148$ \\
$\checkmark$  & $\times$ & $53.244$ & $0.0378$ & $67.272$ & $27.341$ \\
$\times$  & $\checkmark$ & $47.432$ & $0.0229$ & $55.962$ & $23.124$ \\
\rowcolor{pink!30}
$\checkmark$ & $\checkmark$ & $\boldsymbol{43.432}$ & $\boldsymbol{0.0179}$ & $\boldsymbol{50.962}$ & $\boldsymbol{20.124}$ \\
\bottomrule
\end{tabular}
\label{tab:ablation_table}
\end{table}

We quantitatively evaluate state-of-the-art models, with results in Table \ref{tab:overall_table}. The VDM \cite{ho2022video} outperforms GANs due to its progressive denoising with spatial and temporal attention. Conditioning on motion further improves performance, as seen in ControlNet \cite{zhang2023adding}. Our method achieves the best results by leveraging a specialized motion encoder (STME) on the displacement field and training the video diffusion model from scratch rather than fine-tuning.

\begin{figure*}[!t]
\includegraphics[width=\linewidth, trim={0.7cm 0.4cm 0.8cm 0.5cm}, clip]{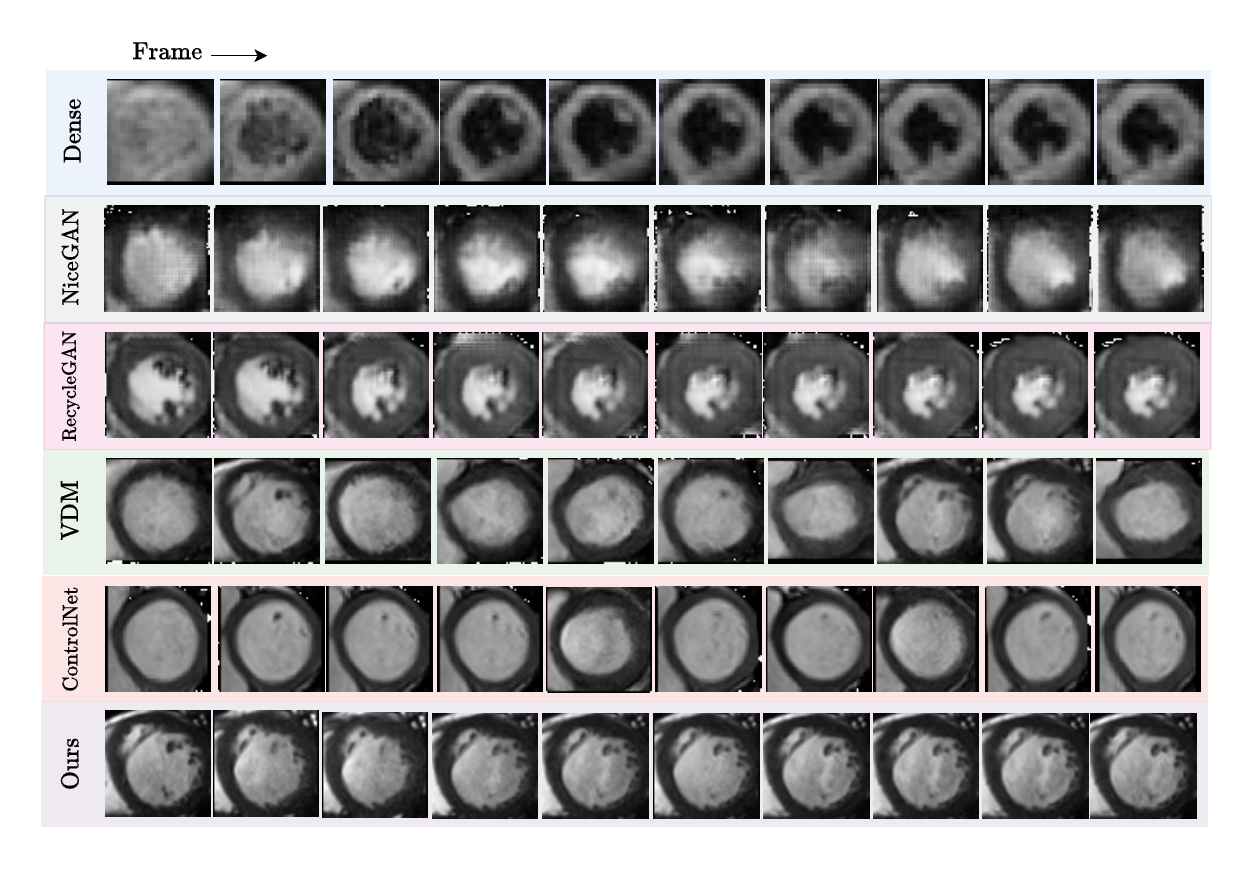}
\caption{Qualitative comparison of Cine CMR synthesis. Our approach (bottom row) preserves both spatial and temporal coherence, while VDM struggles with anatomical consistency during motion, and ControlNet shows improved temporal stability but slightly reduced spatial details.}
\label{fig:qualitative_analysis}
\end{figure*}

We also qualitatively demonstrate the effectiveness of our model in generating realistic, high-contrast cine CMR (see Fig.~\ref{fig:qualitative_analysis}). In the figure, each column represents an individual frame, while each row corresponds to a synthesis sequence from different baseline methods. As shown, our approach produces myocardial sequences that maintain both spatial and temporal coherence across all frames.

We evaluate the effectiveness of our LTA registration network and the STME block in table \ref{tab:ablation_table}. Our experimental finding demonstrate that the absence of LTA causes degradation in the quality of the synthesize cardiac sequences due to lack of spatiotemporal continuity in the motion, while inclusion of LTA further improves the performance by effectively capturing long-range temporal dependencies present within the sequence. Finally, results further improved with the STME block since it enables to extract essential motion characteristics as condition that enhance synthesis quality.

We evaluate our synthesized cine CMR sequences on a downstream segmentation task using a pretrained 3D U-Net \cite{cciccek20163d}. Results show that direct inference on DENSE CMR sequences significantly degrades performance (Dice score: 0.31), whereas our synthesized images improve segmentation accuracy, boosting the Dice score to 0.81 - a 70\% relative increase.

\section{Conclusion}
In this paper, we introduced MFD-V2V, a novel motion feature guided diffusion model for unpaired cardiac video-to-video (V2V) translation, which enables the synthesis of high-contrast cine CMR from low-SNR DENSE CMR sequences. Our approach integrates a latent temporal multihead attention (LTMA) registration network to effectively capture cardiac motion, and a spatiotemporal motion encoder (STME) to enable motion feature guided diffusion model. Through extensive experiments on multi-site cardiac MRI datasets, we demonstrated that MFD-V2V outperforms state-of-the-art V2V generative models in both quantitative evaluation metrics and qualitative assessments, producing more realistic and temporally coherent cardiac MR videos. 

\bibliographystyle{splncs04}
\bibliography{refrences}
\end{document}